\documentclass{ws-procs975x65}
\def\bea{\begin{eqnarray}}
\def\eea{\end{eqnarray}}
\def\bean{\begin{eqnarray*}}
\def\eean{\end{eqnarray*}}

\def\bvec#1{\raise1.5ex\hbox{$\rightarrow$}\mkern-16.5mu #1}
\def\sc#1{\textstyle{\scriptstyle #1}}

\def\m#1{\mathcal#1}
\def\parp{\partial^+}
\newcommand{\be}{\begin{equation}}
\newcommand{\ee}{\end{equation}}

\newcommand{\barr}{\begin{array}}
\newcommand{\earr}{\end{array}}

\newcommand{\bed}{\begin{displaymath}}
\newcommand{\eed}{\end{displaymath}}
\newcommand{\bal}{\begin{array}{ll}}
\newcommand{\eal}{\end{array}}

\newcommand{\deltab}{\boldsymbol\delta}

\renewcommand{\d}{\partial}
\newcommand{\nn}{\nonumber}
\newcommand{\E}{E_{7(7)}}
\newcommand{\ka}{\kappa}
\newcommand{\ep}{\epsilon}

\begin{document}

\title{MAXIMAL SUPERSYMMETRY AND EXCEPTIONAL GROUPS }

\author{LARS BRINK}

\address{Department of Fundamental Physics, Chalmers University of Technology,\\
Gšteborg, Sweden\\
$^*$E-mail: lars.brink@chalmers.se\\
http://fy.chalmers.se/$\sim $tfelb/}

\begin{abstract}
The article is a tribute to my old mentor, collaborator and friend Murray Gell-Mann. In it I describe work by Pierre Ramond, Sung-Soo Kim and myself where we describe the ${\cal N}=8$ Supergravity in the light-cone formalism. We show how the Cremmer-Julia $\E$ non-linear symmetry is implemented  and how the full supermultiplet is a representation of the $\E$ symmetry. I also show how the $\E$ symmetry is a key to understand the higher order couplings in the theory and is very useful when we discuss possible counterterms for this theory.
\end{abstract}

\keywords{Maximal Supersymmetry, Supergravity, Exceptional Symmetries; Proceedings; World Scientific Publishing.}

\bodymatter

\section{Introduction}
Murray Gell-Mann was the leading scientist in particle physics for some twenty years, a very long time in such a competitive subject. When I started as a young graduate student I found his name behind almost all the developments in the field that I studied and when I eventually saw him first at a Nobelsymposium here in Sweden and then at CERN, where I was a fellow in the beginning of the 70's, he made such an impression on me that his stature got even bigger. In 1976 I came to Caltech and when I met him my body was shaking. I was so nervous. After some time we got closer and we started to collaborate. This was my most lucky time in life. All through since then we have been very good friends and our discussion have been on-going even if they sometimes have been interrupted for a year or two. We always made a point to start again exactly where we ended the year before. Murray taught me many things, especially the beauty of symmetries. It has since then been a leading star in my scientific life to follow the symmetries and try to implement as big a symmetry as possible. The possible breaking of the symmetry can come later.

Almost thirty years ago I started to study supersymmetric theories in the light-cone gauge/frame formulation. The starting point was  the ${\cal N}=4$  SuperYang-Mills theory \cite{BSS}. In order to implement as much supersymmetry as possible we \cite{Brink:1982pd} found that if we choose the light-cone gauge $A^+=\frac{1}{\sqrt{2}}(A^0 + A^3) =0$ for the vector fields and choose the light-cone direction $x^+$ as the evolution parameter we could solve for $A^-$. We could also solve  for half the spinors leaving a theory with only the physical degrees of freedom. (The unphysical degrees of freedom satisfies algebraic equations when $x^+$ is the "time" and $x^-$ "space".) They could then be assembled to a superfield and the action could be written in terms of this superfield. With this formalism we could prove that  the ${\cal N}=4$ is perturbatively finite \cite{UVfinite}. We \cite{Bengtsson:1983pg} also constructed the  ${\cal N}=8$ Supergravity \cite{Cremmer:1979up} this way but failed to construct higher order couplings than the three-point ones.

Some years ago Pierre Ramond and I took up this programme again in the light of the modern developments.  The maximally supersymmetric ${\cal N}=8$ Supergravity and ${\cal N}=4$ SuperYang-Mills  play a very important r\^ole in modern theory. In the standard descriptions they look quite different and are naturally related to eleven- and ten-dimensional theories, respectively. In the light-cone frame description, however, they are described in a remarkably similar way hinting at a deep relation between them. In four dimension,  they are the only two (except possibly for some higher-spin theories) that are  described by one chiral {\em constrained} light-cone superfield which captures {\em all} their physical degrees of freedom \cite{Brink:1982pd}.  Also,  tree level Supergravity amplitudes are related to the square of Yang-Mills amplitudes \cite{Berends:1988zp, Kawai:1985xq}, and the light-cone Hamiltonian of both theories can be written as a {\em positive definite} quadratic form in their superfields \cite{Ananth:2005zd, Ananth:2006fh}. Some even suggest that the ultraviolet finiteness of  ${\cal N}=4$ SuperYang-Mills \cite{UVfinite} might extend to ${\cal N}=8$ Supergravity \cite{Bern:1998ug,Green:2006gt}. There are important structural differences though.  ${\cal N}=8$ Supergravity, unlike ${\cal N}=4$ SuperYang-Mills, is not Superconformal invariant. Instead it has  the on-shell, non-linear Cremmer-Julia,  $\E$ duality symmetry \cite{Cremmer:1979up}. It was therefore natural to ask if this symmetry can be exploited to bring simplicty to the quartic and higher-order interactions of $\m N=8$ Supergravity. This was done in a paper that we wrote some time ago \cite{Brink:2008qc}, which I will describe here.

There are two approaches that one can follow to find the Hamiltonian or the action in this formalism. On the one hand we can follow the scheme mentioned above by fixing the light-cone gauge and eliminate all unphysical degrees of freedom. The other one is to look for a (non-linear) representation of the SuperPoincar\'e based on the superfield. In our work we have had to use both methods as I will indicate later.

\section{$\m N=8$ Supergravity in Light-Cone Superspace}
The $256$ physical degrees of freedom of ${\cal N}=8$ Supergravity form {\em one} constrained chiral superfield in the superspace spanned by eight Grassmann variables, $\theta^m$ and their complex conjugates $\bar\theta_m$ $(m=1,...,8)$, on which $SU(8)$ acts linearly. 
We introduce the chiral derivatives

\bea
d^m ~\equiv~ -\frac{\d}{\d\bar\theta_m} -\frac{i}{\sqrt{2}} \theta^m\d^+ \, , ~~~ 
\bar d_m ~\equiv~ \frac{\d}{\d\theta^m} +\frac{i}{\sqrt{2}} \bar\theta_m\d^+ \ , 
\eea
which satisfy canonical anticommutation relations

\begin{equation}\label{phianticomm}
\left\{ d^m\,,\,\bar d_n\right\}~=~ -i \sqrt{2}\delta^m{}_{n} \d^+ \ .
\end{equation}
The physical degrees of freedom of  ${\cal N}=8$ Supergravity, the spin-2 graviton $h$ and $\overline h$;  eight spin-$\frac{3}{2}$ gravitinos, ${\psi}^m$ and 
${\overline \psi}_m$, twenty-eight vector fields 

\[
 \overline B_{mn} \equiv \frac{1}{\sqrt{2}} \left( B^1_{mn} \,+\,i\,B^2_{mn}\right)\ ,
 \]
and their conjugates, fifty-six  gauginos ${\overline \chi}_{mnp}$ and $\chi^{mnp}$, and finally seventy real scalars ${\overline D}_{mnpq}$ appear in one superfield

\begin{eqnarray}\label{superfield}
\varphi\,(\,y\,)\,&=&\,\frac{1}{{\d^+}^2}\,h\,(y)\,+\,i\,\theta^m\,\frac{1}{{\d^+}^2}\,{\overline \psi}_m\,(y)\,+\,i\,\theta^{mn}_{}\,\frac{1}{\d^+}\,{\overline B}_{mn}\,(y)\nn\  \\
\;&&-\,\theta^{mnp}_{}\,\frac{1}{\d^+}\,{\overline \chi}^{}_{mnp}\,(y)\,-\,\theta^{mnpq}_{}\,{\overline D}^{}_{mnpq}\,(y)+\,i\widetilde\theta^{}_{~mnp}\,\chi^{mnp}\,(y)\nn \\
&&+\,i\widetilde\theta^{}_{~mn}\,\d^+\,B^{mn}\,(y)+\,\widetilde\theta^{}_{~m}\,\d^+\,\psi^m_{}\,(y)+\,{4}\,\widetilde\theta\,{\d^+}^2\,{\bar h}\,(y)\ ,
\end{eqnarray}
where the bar denotes complex conjugation, and 
\[
\theta^{a_1a_2...a_n}_{}~=~\frac{1}{n!}\,\theta^{a_1}\theta^{a_2}_{}\cdots\theta^{a_n}_{}\ ,\quad \widetilde\theta^{}_{~a_1a_2...a_{n}}~=~ \epsilon^{}_{a_1a_2...a_{n}b_1b_2...b_{(8-n)}}\,\theta_{}^{b_1b_2\cdots b_{(8-n)}}\,\ .
\]
The arguments of the fields are the chiral coordinates 
\[
y~=~(x,\,\bar x,\, x^+, \,y^-\equiv x^- -\frac{i}{\sqrt2}\theta^m\bar\theta_m\, )\ ,\qquad  x=\frac{1}{\sqrt{2}}(x_1+ix_2)\ ,
\]
so that  $\varphi$ and its complex conjugate $\overline \varphi$ satisfy the chiral constraints

\bea\label{chiralconstraints}
d_{}^m\, \varphi ~=~0, \qquad \overline d^{}_m\, \overline \varphi ~=~0\ ,
\eea
The complex chiral superfield is related to its complex conjugate by the {\em inside-out constraint}

\begin{equation}\label{insideout}
\varphi~=~\frac{1}{4\,\d^{+4}}\,d_{}^1d_{}^2\cdots d_{}^8\, \overline\varphi\ ,
\end{equation}
in accordance with the duality condition of ${ D}^{mnpq}$. 

\[
D^{mnpq} =\frac{1}{4!} \epsilon^{mnpqrstu} \overline{D}_{rstu}.
\]

 On the light-cone, the eight kinematical supersymmetries (the spectrum-generating part of the symmetry) are linearly represented 
by the operators $q^m$  and $\bar q_m$. (We use the light-cone notation for the generators of the SuperPoincar\'e algebra, dividing the spinors into two two-components spinors, which are then linearly combined into complex anticommuting ones.) 

\begin{equation}
q^m ~=~ -\frac{\d}{\d\bar\theta_m} +\frac{i}{\sqrt{2}} \theta^m\d^+ \, , ~~~ 
\bar q_m ~=~ \frac{\d}{\d\theta^m} -\frac{i}{\sqrt{2}} \bar\theta_m\d^+ \ ,
\end{equation}
which also satisfy anticommutation relation

\begin{equation}
\{\, q^m\, , \, \bar q_n\,\} ~=~ i\sqrt{2}\, \delta^m{}_n \,\d^+\ ,
\end{equation}
and anticommute with the chiral derivatives. Hence, their linear action on the chiral superfield 

\bea
\delta^{}_s\,\varphi(y)~=~\overline \epsilon_m^{}\,q_{}^m\,\varphi(y)\ ,
\eea
where $\bar\epsilon_m$ is the parameter of the supersymmetry transformation, preserves chirality. The kinematical supersymmetry transformations of the physical fields are then 

\[
\delta^{}_s\,h~=~0\ ,\qquad \delta^{}_s\,\overline h~=~-i\frac{\sqrt{2}}{4}\,\overline\epsilon_m\,\psi^m_{}\ ,\]
\[
\delta^{}_s\,\psi^m_{}~=~2\sqrt{2}\,\overline\epsilon_n\partial^+\,B^{mn}_{}\ ,\qquad \delta^{}_s\,\overline\psi^{}_m~=~-\sqrt{2}\,\overline\epsilon_m\,\partial^+\,h\ ,\]
\[
\delta^{}_s\,B^{mn}_{}~=~-3i\sqrt{2}\,\overline\epsilon_p\,\chi^{mnp}_{}\ ,\qquad \delta^{}_s\,\overline B^{}_{mn}~=~-2i\sqrt{2}\,\overline\epsilon_{[m}\overline\psi^{}_{n]}\ ,\]
\[
\delta^{}_s\,\chi^{lmn}_{}~=~-\frac{\sqrt{2}}{3!}\,\overline\epsilon_k\,\partial^+D^{klmn}_{}\ ,\qquad \delta^{}_s\,\overline\chi^{}_{mnp}~=~-3\sqrt{2}\,\epsilon^{}_{[p}\,\partial^+\,\overline B^{}_{mn]}\ ,\]
and finally

\[
\delta^{}_s\,\overline D^{}_{klmn}~=~-4i\sqrt{2}\,\overline\epsilon^{}_{[n}\,\overline\chi^{}_{klm]}\ .\]

The quadratic operators 

\begin{equation}\label{su8fromq}
T^i{}_j ~=~ -\frac{i}{\sqrt{2} \,\d^+} \left( q^i \bar q_j -\frac{1}{8}\delta^i{}_j q^k \bar q_k \right) \ ,
\end{equation}
which satisfy the  $SU(8)$ algebra

\[
\  [\,T^i{}_j\,,\,T^k{}_l\,] ~=~\delta^k{}_j\, T^i{}_l - \delta^i{}_l \,T^k{}_j \ ,
\]
also act linearly on the chiral superfield 

\[
 \delta^{}_{SU_8} \, \varphi(y)~=~ \omega^j{}_{i}\,T^{i}{}_{j}\, \varphi(y)\ .
 \]
They constitute the $R$-symmetry of the theory. Similarly we have found  the linear part of the representation of the remaining generators of the SuperPoincar\'e algebra. All generators with a minus-component are generators that take the field forward in the light-cone time and they get non-linear contributions in the interacting theory. All the three-point couplings were found already in the paper \cite{Bengtsson:1983pg}. These {\it dynamical} generators include $P^- = H$, the Hamiltonian, where we also used the masslessness condtion $P^2=0$ and  the other part of the supersymmetry $Q_-$. In our work on  the ${\cal N}=4$  SuperYang-Mills theory \cite{Ananth:2005zd} we found that maximally supersymmetric theories have the unique property that 

\bea
H_{}~=~\frac{1}{2\,\sqrt{2}}\,\int{d^4}x\,{d^4}\theta\,{d^4}{\bar \theta}\,\,{\bar {\mathcal W}}\,\cdot\,{\mathcal W}\, ,
\eea
where

\be
W = \delta^{}_{Q_-} \, \varphi
\ee
and $\cdot$ represents a power of $\frac{1}{\partial^+}$ to make the integral dimensionally correct. This fact was used for the ${\cal N}=8$ Supergravity Theory in ref. (9) to find the four-point coupling by a trial and error method. It was found to consist of some 96 terms with no obvious relations among them. It was clear that we needed some further insight.

\section{The  $\E$ Symmetry in the Light-Cone Formulation}

The  $\E$ symmetry has been quite instrumental in the construction and the interpretation of the covariant formulation of the  ${\cal N}=8$ Supergravity Theory \cite{Cremmer:1979up}. In this formulation it is a symmetry at the level of the equations of motion and it only affects the vector and the scalar particles. In our light-cone formulation it must act on the whole superfield. One could try to find a non-linear representation on the whole superfield but the method we eventually found to work was to go back to the original works \cite{Cremmer:1979up, bernard} and implement the light-cone gauge condition on the vector field and then eliminate all unphysical degrees of freedom by solving their algebraic equations of motion. We could so eliminate time-derivatives in the action by non-linear field redefinitions and read off the $\E/SU(8)$ transformations of the new vector and scalar fields. However, we then found that these transformations do not close properly. All the fields have to transform as I alluded to above. We could include the other fields of the theory by demanding that the $\E/SU(8)$ transformations commute with the kinematical supersymmetries, that is

\bea
[\,\delta^{}_q\,,\,\delta^{}_{E/SU_8}]\,\varphi~=~0 \ .
\eea

In this way we could read off the transformations for all the fields and check that the transformations close properly.
 Including the inhomogeneous term, the $\E/SU(8)$ transformation could be written in a compact way by introducing a coherent state-like representation 

\begin{equation}\label{}
\delta^{}_{E/SU_8}\,\varphi~=~
-\frac{2}{\kappa}\,\theta^{ijkl}_{}\,\overline\Xi^{}_{ijkl}\,+\,
\frac{\kappa}{4!}\,\Xi^{ijkl}  \left(\frac{\d}{\d\eta}\right)_{ijkl}\frac{1}{\partial^{+2}}\left(e^{\eta \hat{\bar d}} \d^{+3} \varphi\, e^{-\eta \hat{\bar d}}\d^{+3} \varphi \right)\Bigg|_{\eta=0}\,+\, \m O(\ka^3)\ ,
\end{equation}
where 

$$
\eta\hat{\bar d} = \eta^m\frac{\bar d_m}{\d^+},~~{\rm and}~~\left(\frac{\d}{\d\eta}\right)_{ijkl} \equiv~ \frac{\d}{\d\eta^i}\frac{\d}{\d\eta^j}\frac{\d}{\d\eta^k}\frac{\d}{\d\eta^l}\ .
$$
We note that these $\E/SU(8)$ transformations do close properly to an $SU(8)$ transformation on the superfield

$$ [\,{ \delta^{}_{E/SU_8}}_1\,,\,{\delta^{}_{E/SU_8}}_2\,] \, \varphi~=~ \delta_{SU(8)}\, \varphi\ .$$ 
It is chiral by construction $d^n_{}\delta^{}_{E/SU_8}\varphi~=~0$, with the power of the first inverse derivative set by comparing with the graviton transformation. 
Hence, {\em all} physical fields, including the graviton transform under $\E$ and can be read off from this equation. I have left out all details of these calculations and I refer the reader to the paper \cite{Brink:2008qc} for all the details.

We can now extend the method to the dynamical supersymmetries, and determine the form of the interactions implied by the $\E$ symmetry.

%%%%%%%%%%%%%%%%%%%%%%%%%%%%%%%%%%%%%%%%%%%%%%
\subsection{ Superspace Action}
{The ${\cal N}=8$ Supergravity action in superspace was first obtained in \cite{Brink:1979nt} and its $LC_2$ form is derived in  \cite{Bengtsson:1983pg} to order $\kappa$,} using algebraic consistency and simplified further in \cite{Ananth:2005vg}. It is remarkably simple:

\begin{equation}\label{superfieldaction}
S = -\frac{1}{64}\int d^4x\,\int d^8\theta\,d^8\overline\theta \,\,\left\{\,-\,\overline\varphi\,\frac{\Box}{\d_{}^{+4}}\,\varphi
\,-\,2\kappa \left(\frac{1}{\d^{+2}}\overline\varphi
\,\overline\d\varphi\,\overline\d\varphi\,+\, c.c.\right)\,+\, \m O(\ka^2) \right\}\ ,
\end{equation}
where $\Box\equiv 2 \,(\,\d\bar\d\,- \,\-\d^+\d^- )$.
The light-cone superfield Hamiltonian density is then written as

\begin{eqnarray}\label{}
{\cal H}~=~2\,\overline\varphi\,\frac{\d\bar\d}{\d^{+4}}\,\varphi
\,+\,2\, \kappa \left(\frac{1}{\d^{+2}}\overline\varphi
\,\overline\d\varphi\,\overline\d\varphi\,+\, c.c.\right)\,+\, {\cal O}(\kappa^2)\ .
\end{eqnarray}
It can be derived from the action of the dynamical supersymmetries on the chiral superfield

\begin{eqnarray}\label{dyn-susy}
\delta^{dyn}_s\,\varphi&=&\delta^{dyn\,(0)}_s\,\varphi\,+\,\delta^{dyn\,(1)}_s\,\varphi\,+\,\delta^{dyn\,(2)}_s\,\varphi\,+\,\m O(\ka^3)\ ,\\
&=&\epsilon^m \left\{\frac{\partial}{\partial^+}\,\bar q_m\,\varphi+\,\kappa\,\frac{1}{\parp}\,{\Big (}\,{\bar \partial}\,{\bar d}_m\,\varphi\,{\d^{+2}}\,\varphi\,-\,\parp\,{\bar d}_m\,\varphi\,\parp\,{\bar \partial}\,\varphi\,
{\Big )}\,+\m O(\kappa^2)\right\}\ .\nn
\end{eqnarray}

We now require that the $\E/SU(8)$ commutes with the dynamical supersymmetries. Let me introduce the notation $\delta_{\E/SU(8)} = \deltab$.

\begin{equation}\label{4.18}
[\,\deltab\,,\,\delta^{dyn}_s\,]\,\varphi~=~0\ .
\end{equation}
This commutativity is valid only on the chiral superfield. 
For example, $[\,\deltab_1\,,\, \delta_{s}\,]\,\deltab_2 \varphi \,\ne\,0$, due to the non-linearity of the $\E$ transformation.
% commutativity with the supersymmetry on the chiral superfield (\ref{4.18}) , does not yield $[\,\delta{\E/SU(8)}\,,\, \delta_{s}\,]\, \deltab \varphi\,=\,0 $. 
 This helps us understand how the Jacobi identity
 
$$ \left(\,  [\, \deltab_1\,,\,[\,\deltab_2\,,\, \delta_{s}\,]\,]  \,+\, [\, \deltab_2\,,\,[\,\delta_{s}\,,\, \deltab_1\,]\,] \, +\, [\, \delta_{s}\,,\,[\,\deltab_1\,,\, \deltab_2\,]\,] \,\right)\, \varphi~ =~0\ ,$$ 
is algebraically consistent. In the last  term the commutator of the  two $\E/SU(8)$ transformations,  $[\,\deltab_1\,,\, \deltab_2\,]$, yields an  $SU(8)$ under which the supersymmetry transforms. This is precisely compensated by contributions from the first two terms.

Although the dynamical supersymmetry to order $\kappa$ is already known, we re-derive $\delta^{dyn\,(1)}_s\,\varphi$ from the commutativity between the dynamical supersymmetries and $\E/SU(8)$ transformations. 

The inhomogeneous $\E$ transformations link interaction terms with different order in $\kappa$. To zeroth order, one finds

\begin{equation}\label{70swith-susy-tofirstorder}
\ [ \, \boldsymbol\delta^{(-1)}\,,\,\delta^{dyn\,(1)}_{s} \, ] \, \varphi  ~=~\deltab^{(-1)}\,\delta^{dyn\,(1)}_{s} \varphi~=~0 \ ,
\end{equation}
since $\delta^{dyn\,(1)}_{s}\,\deltab^{(-1)} \varphi~=~0$. To find $\delta^{dyn\,(1)}_{s} \varphi$ that satisfies both the above equation and the SuperPoincar\'e algebra, one may start with a general form that satisfies all the commutation relations with the kinematical SuperPoincar\'e generators (the forms of the kinematical SuperPoincar\'e generators can be found in \cite{Bengtsson:1983}),

$$ \delta^{dyn\,(1)}_{s} \varphi \propto \frac{\d}{\d a}\frac{\d}{\d b}\,\frac{1}{\d^{+(m+n+1)}} \left( e^{a \hat{\bar \d}}e^{b\,\ep \hat{\bar q}} \d^{+(2+m)}\varphi\, e^{-a \hat{\bar \d}}e^{-b\,\ep \hat{\bar q}}\d^{+(2+n)}\varphi \right) \Big|_{ a=b=0 }\ ,$$ 
where $\hat{\bar \d} = \frac{\bar\d}{\d^+}$, $\ep\hat{\bar q} = \ep^m\frac{\bar q_m}{\d^+}$.
It is not difficult to see that this form with non-negative $m,\, n$ satisfies (\ref{70swith-susy-tofirstorder}). The number of powers of $\d^+$ can be determined by checking the commutation relation between two dynamical generators $\delta_{p^-}$(Hamiltonian variation which is derived from the supersymmetry algebra) and $\delta^{}_{j^-}$(the boost which can also be obtained through $ [\,\delta^{}_{j^-}\,,\, \delta_{\bar q} \,]\,\varphi\, =\, \delta^{dyn}_{s}\varphi$), yielding that the commutator between $\delta^{}_{j^-}$ and $\delta_{p^-}$ vanishes only when $m=n=0$,
which leads to the the same form as (\ref{dyn-susy}) written in a coherent-like form

$$
\delta^{dyn\,(1)}_{s} \varphi = \frac{\ka}{2} \frac{\d}{\d a}\frac{\d}{\d b}\,\frac{1}{\d^{+}} \left[ e^{a \hat{\bar \d}}e^{b\,\ep \hat{\bar q}} \d^{+2}\varphi\, e^{-a \hat{\bar \d}}e^{-b\,\ep \hat{\bar q}}\d^{+2}\varphi \right] \Big|_{a=b=0 }\ .$$
It is worth noting that this is the solution that has the least number of powers of $\d^{+}$ in the denominator, and thus the least ``non-local''. 

The same reasoning can be applied to higher orders in $\kappa$.  To order $\ka$, we find that commutativity

\be\label{susywithe} \ [ \, \boldsymbol\delta^{(-1)}\,,\,\delta^{dyn\,(2)}_{s} \, ] \, \varphi \, +\,[ \, \boldsymbol\delta^{(1)}\,,\,\delta^{dyn\,(0)}_{s} \, ] \, \varphi ~=~0\ \ee
requires

\begin{eqnarray}\label{delta2qvarphi1}
&&\deltab^{(-1)}\,\delta^{dyn\,(2)}_{s} \varphi\\
&&=~\frac{\kappa}{4!} \Xi^{ijkl} \frac{1}{\partial^{+3}} \left[ \,
-\,{\bar d}^{}_{ijkl}\frac{\partial}{\partial^+} \varphi \, \partial^{+3}\epsilon\bar q \,\varphi
+\,4\, {\bar d}^{}_{ijk} \partial \varphi \, \bar d_l \partial^{+2} {\epsilon\bar q}\, \varphi 
-\, 3\,{\bar d}^{}_{ij}\partial \partial^+ \varphi\, {\bar d}^{}_{kl}\partial^+{\epsilon\bar q}\,\varphi \right.\nn\\
&&\qquad\qquad\qquad~~
~ - \,{\bar{d}}^{}_{ijkl} \frac{\epsilon\bar q}{\partial^+}\, \varphi \,\partial\partial^{+3}\varphi
+\, 4\, {\bar d}^{}_{ijk} {\epsilon\bar q}\, \varphi \, \bar d_l \partial \partial^{+2}\,\varphi  
-3 \,{\bar d}^{}_{ij} \partial^{+} {\epsilon\bar q}\varphi\, {\bar d}^{}_{kl}\partial\partial^+\varphi \nn\\
%%%
&&\qquad\qquad\qquad~~
~ +\, {\bar d}^{}_{ijkl}\frac{\partial}{\partial^{+2}} \epsilon\bar q\,\varphi \, \partial^{+4}\varphi
\,-\, 4\, {\bar d}^{}_{ijk} \frac{\partial}{\partial^+} {\epsilon\bar q} \,\varphi \, \bar d_l \partial^{+3}\,\varphi  
\,+\,  3\,{\bar d}^{}_{ij}\partial {\epsilon\bar q}\,\varphi\, {\bar d}^{}_{kl}\partial^{+2}\varphi \nn\\
%%%%
&&\qquad\qquad\qquad~~
~+\,{\bar d}^{}_{ijkl}\,\varphi \, \partial \partial^{+2}{\epsilon\bar q}\, \varphi 
\,-\,4\, {\bar d}^{}_{ijk} \partial^+\, \varphi \, \bar d_l \partial \partial^{+}{\epsilon\bar q}\,\varphi  
\,+\,  3\,{\bar d}^{}_{ij} \partial^{+2} \varphi\, {\bar d}^{}_{kl}\partial{\epsilon\bar q}\varphi \bigg]\ ,\nn
\end{eqnarray}
where $\epsilon \bar q$ denotes $\epsilon^m\bar q_m$, which can be written in a simpler form by rewriting it in terms of a coherent state-like form: 

\begin{eqnarray}\label{coherent-state-form}
&&\deltab^{(-1)}_{}\,\delta^{dyn\,(2)}_{s} \varphi \\
&&=~  
 \frac{\kappa}{2\cdot4!} \Xi^{ijkl} \frac{\d}{\d a}\frac{\d}{\d b}  
 \left(\frac{\d}{\d\eta}\right)_{ijkl}
 \frac{1}{\partial^{+3}} \left[\
 e^{a \hat{\d}} e^{b\, \epsilon\hat{\bar q}}e^{\eta\hat{\bar d}}\d^{+4} \varphi\,\, e^{-a\hat{\d}}e^{-b \,\epsilon\hat{ \bar q}} e^{-\eta\hat{\bar d}}\d^{+4} \varphi
\, \right] \Bigg|_{a=b=\eta=0}\ .\nn
\end{eqnarray}

To find $\delta^{dyn (2)}_{s}\varphi$ that satisfies (\ref{delta2qvarphi1}), consider the chiral combination

\begin{eqnarray}\label{}
Z_{mnpq}&\equiv&\left(\frac{\d}{\d\xi}\right)_{mnpq}\left( e^{\xi\hat{\bar d}}  \d^{+4}\varphi e^{-\xi\hat{\bar d}}  \d^{+4}\varphi \right) \Big|_{\xi =0}\ ,\\
&=&
\bar d^{}_{mnpq} \varphi\, \partial^{+4} \varphi \,-\,4\,\bar d^{}_{mnp}\partial^+\varphi \,\bar d_q \partial^{+3} \varphi \,+\,3\,\bar d^{}_{mn}\partial^{+2}\varphi \,\bar d^{}_{pq} \partial^{+2} \varphi \ .\nn
\end{eqnarray}
The inhomogeneous $\E$ transformation of

$$Z^{ijkl}\equiv\frac{1}{4!}\epsilon^{ijklmnpq}Z_{mnpq}\ ,$$
has the simple form 
\bea\label{}
\boldsymbol\delta^{(-1)} Z^{ijkl} =\frac{1}{4!}\epsilon^{ijklmnpq}\,\bar d^{}_{mnpq} \boldsymbol\delta^{(-1)} \varphi\, \partial^{+4} \varphi =
\frac{2}{\kappa} \,\Xi^{ijkl}\, \partial^{+4} \varphi \ ,
\eea
which leads to the solution

\be\label{s2}
\delta^{dyn\,(2)}_s\,\varphi~=~
\frac{\kappa^2}{2\cdot4!}\frac{\d}{\d a}\frac{\d}{\d b}\left(\frac{\d}{\d\eta}\right)_{ijkl}
\frac{1}{\d^{+4}}\left(\,
 e^{a \hat{\d}+b\, \epsilon\hat{\bar q}+\eta\hat{\bar d}}\d^{+5} \varphi\,\,
 e^{-a \hat{\d}-b\,\epsilon \hat{\bar q}-\eta\hat{\bar d}}  Z^{ijkl}
\right)
\, \Bigg|_{a=b=\eta=0}\ ,
\ee
where we have fixed the ambiguity discussed earlier by choosing the expression with the least number of $\d^+$ in the denominator.  This coherent state-like form is very efficient; Written out explicitly $\delta^{dyn\,(2)}_s\varphi$ consists of 60 terms.

The dynamical supersymmetry is then written in terms of the coherent state-like form, 

\begin{eqnarray}\label{solq2}
\delta^{dyn}_s\,\varphi
&=& \frac{\d}{\d a}\frac{\d}{\d b} \Bigg\{ e^{a\hat{\d}} e^{b\,\ep \hat{\bar q}} \d^+\varphi \, +\,
\frac{\kappa}{2}\,\frac{1}{\d^{+}}\left(e^{a\hat{\bar \d}+b\,\ep\hat{\bar q}} \d^{+2}\varphi e^{-a\hat{\bar \d}-b\,\ep\hat{\bar q}} \d^{+2}\varphi \right) \nn\\
&&~+\,\frac{\kappa^2}{2\cdot4!}\left(\frac{\d}{\d\eta}\right)_{ijkl}
\frac{1}{\d^{+4}}\left(\,
 e^{a \hat{\d}+\, b\,\epsilon\hat{\bar q}+\eta\hat{\bar d}}\d^{+5} \varphi\,\,
 e^{-a \hat{\d}-\,b\,\epsilon \hat{\bar q}-\eta\hat{\bar d}}  Z^{ijkl}
\right)  \nn \\
&&~+\,\m O(\ka^3) \Bigg\}
 \Bigg|_{a=b=\eta=0} \ .
\end{eqnarray}

%%%%
We now use the fact, as  Ananth et al \cite{Ananth:2006fh} have shown, that the $\m N =8$ supergravity light-cone Hamiltonian can be written as a quadratic form (to order $\kappa^2$), 

\[ \m 
H~= ~ \frac{1}{4\sqrt2}\, \left( \m W_m\,,\, \m W_m\right)~\equiv~\frac{2\,i}{4\sqrt2} \int  d^8 \theta\,d^8\bar\theta\,d^4x\,\,  \overline{\m W}_m\frac{1}{\d^{+3}}\m W_m \ ,  \]
where the fermionic superfield $\m W_m$ is  the dynamical supersymmetry variation of $\varphi$ 

\[
\delta^{dyn}_{s} \varphi~\equiv~ \epsilon^m_{}\,\m W^{}_m\ ,\]
with 

\[\m W_m~=~ \m W^{(0)}_{m} \, +\,\,\m W^{(1)}_{m} \, +\,\,\m W^{(2)}_{m} +\,\cdots\ . \]
Up to order $\ka$, the Hamiltonian is simply

\begin{equation}\label{}
\m H~=~ \frac{1}{4\sqrt2}\, \left[ \left( \m W^{(0)}_m\,,\, \m W^{(0)}_m\right)\,+ \,\left( \m W^{(0)}_m\,,\, \m W^{(1)}_m\right)\,+ \, \left( \m W^{(1)}_m\,,\, \m W^{(0)}_m\right)\right] \ ,
\end{equation}
while the Hamiltonian of order $\ka^2$ consists of three parts:

\begin{equation}\label{}
 \m H^{\ka^2} ~=~ \frac{1}{4\sqrt2}\, \left[  \left( \m W^{(1)}_m\,,\, \m W^{(1)}_m\right)\,+\,\left( \m W^{(0)}_m\,,\, \m W^{(2)}_m\right)\,+ \, \left( \m W^{(2)}_m\,,\, \m W^{(0)}_m\right)\right] \ , 
\end{equation}
where the first part was computed by Ananth et al \cite{Ananth:2006fh}

\begin{eqnarray}\label{}
&&\left( \m W^{(1)}_m\,,\, \m W^{(1)}_m\right)~=~ i\frac{\ka^2}{2}\frac{\d}{\d a}\frac{\d}{\d b}\frac{\d}{\d r}\frac{\d}{\d s} \int d^8 \theta\,d^8\bar\theta\,d^4x\,\,\\
&&
\frac{1}{\d^{+5}} \left(e^{a\hat{\d}+b\hat{q}^m} \d^{+2}\overline\varphi e^{-a\hat{\d}-b\hat{q}^m} \d^{+2}\overline\varphi \right)
\left(e^{r\hat{\bar \d}+s\hat{\bar q}_m} \d^{+2}\varphi e^{-r\hat{\bar \d}-s\hat{\bar q}_m} \d^{+2}\varphi \right)\Big|_{a=b=r=s=0}\ ,\qquad \nn
\end{eqnarray}
 and the second and  third parts are complex conjugate of each other.  It suffices to consider  

\begin{eqnarray}\label{}
 \left( \m W^{(0)}_m\,,\, \m W^{(2)}_m\right)&=& i\frac{\ka^2}{4!}\frac{\d}{\d a}\frac{\d}{\d b}\left(\frac{\d}{\d\eta}\right)_{ijkl}\int d^8 \theta\,d^8\bar\theta\,d^4x\,\,\\
&&~\frac{\bar \d}{\d^+} q^m \overline \varphi\, 
\frac{1}{\partial^{+7}}\left(\,
 e^{a \hat{\d}+b\, \hat{\bar q}_m+\eta\hat{\bar d}}\d^{+5} \varphi\,\,
 e^{-a \hat{\d}-b\, \hat{\bar q}_m-\eta\hat{\bar d}}  Z^{ijkl}
\right) \Bigg|_{a=b=\eta=0} \ .\nn
\end{eqnarray}
Integration by parts with respect to $\bar d$'s and use of the inside-out constraint (\ref{insideout}) allow for  an efficient rearrangement of terms to yield the final expression

\begin{eqnarray}\label{}
&&\left( \m W^{(0)}_m\,,\, \m W^{(2)}_m\right) \\
&&=~-i
\frac{\ka^2}{4!} \frac{\d}{\d a}\frac{\d}{\d b} \int d^8 \theta\,d^8\bar\theta\,d^4x\,\frac{\bar \d}{\d^{+4}} q^m {d}^{ijkl}\overline \varphi
\left(  e^{a \hat{\d}\, +\, b\, \hat{\bar q}_m} \d^+\overline\varphi \,\,e^{-a \hat{\d}\, -\, b \hat{\bar q}_m}
\frac{1}{\d^{+4}} Z_{ijkl}
\right)\Bigg|_{a=b=0} \ .\nn
\end{eqnarray}
Therefore, the Hamiltonian to order $\kappa^2$  is written as

\begin{eqnarray}\label{}
 &&\m H^{\ka^2} ~=~i\,\frac{\ka^2}{4\sqrt2} \int d^8 \theta\,d^8\bar\theta\,d^4x\,\,\frac{\d}{\d a}\frac{\d}{\d b}\\
&&\quad\Bigg\{\frac{1}{2} \frac{\d}{\d r}\frac{\d}{\d s} \frac{1}{\d^{+5}} \left(e^{a\hat{\d}+b\hat{q}} \d^{+2}\overline\varphi e^{-a\hat{\d}-b\hat{q}} \d^{+2}\overline\varphi \right)\left(e^{r\hat{\bar \d}+s\hat{\bar q}} \d^{+2}\varphi e^{-r\hat{\bar \d}-s\hat{\bar q}} \d^{+2}\varphi \right)\nn\\
&&~~~~-\, \left[\,\frac{1}{4!}\, \frac{\bar \d}{\d^{+4}} q^m {d}^{ijkl}\overline \varphi
\left(  e^{a \hat{\d}\, +\, b\, \hat{\bar q}_m} \d^+\overline\varphi \,\,e^{-\,a\, \hat{\d} \,-\, b\, \hat{\bar q}_m}
\frac{1}{\d^{+4}} Z_{ijkl}
\right) +c.c. \right] \Bigg\} \Bigg|_{a=b=r=s=0} \ .\nn
\end{eqnarray}
to be compared with the 96 terms of Ananth et al \cite{Ananth:2006fh}! 

\section{Higher-Point Terms and Possible Counterterms}

Our technique with coherent-state like expressions has given us hope that we could find also higher-point terms in the expansion in the coupling constant $\kappa$. In principle we need to find these terms for all dynamical generators as well as for the $\E$ transformations. The key calculations are extensions of (\ref{susywithe}). There is a good chance that we would find systematics enough to find the higher order terms for both $\E$ symmetry and well as for the dynamical supersymmetry transformations. We have not succeeded in doing this so far and there is one problem that we find more urgent and interesting. That is the following.

In the construction of the  ${\cal N}=4$  SuperYang-Mills theory we could easily prove that the interaction terms in the dynamical generators are unique since the coupling constant is dimensionless and the algebra gives a unique choice for the number of derivatives in the expressions. For a gravity theory the situation is different. The coupling constant  $\kappa$ is dimensionfull and there is a possibility that a higher order of  $\kappa$ could be compensated by more derivatives in the expressions for the generators. In fact it can be shown in ordinary gravity that a possible two-loop counterterm exists \cite{BK}, which of course is a well-known result. Taking over this expression to  ${\cal N}=8$ Supergravity by finding  a superspace expression that contains the gravity result to order $\theta^0$, we found that it should be of the form

\be
\delta_{P^-} \varphi \sim \bar \varphi  \bar \varphi
\ee
This expression which contains six derivatives is evidently not satisfying the chiral constraint and is hence not consistent. The result that there is no three-point counterterm in supergravity has been known since the first year of supergravity \cite{Deser:1977nt}.

Hence the first possible counterterm must be looked for at the four-point level. Remember the condition (\ref{susywithe}). It says that the kinetic term leads uniquely to a four-point coupling with the same number of derivatives as the kinetic one. If there is another possible four-point coupling it must hence satisfy

\be 
 \ [ \, \boldsymbol\delta^{(-1)}\,,\,\delta^{dyn\,(2)}_{s,ct} \, ] \, \varphi \,  ~=~0,\ \ee
where $\delta^{dyn\,(2)}_{s,ct}$ represents a possible counterterm for the dynamical supersymmetry, ie. a term with an even number higher than one of derivatives. This term must then transform correctly under the remaining generators of the SuperPoincar\'e algebra, and the dynamical ones have to be given higher counterterms  too.

We can use the expression (\ref{s2}) as the starting point extending it with exponentials also in $\hat {\bar \d}$ demanding one more $\hat {\d}$ than the number of  $\hat {\bar \d}$'s. We know that it satisfies the correct commutation relations with the kinematic generators. What remains  to be performed are the commutations with the dynamical generators extended to the same order in derivatives and $\kappa$. This work is in progress \cite{BK}. If we could not find any counterterms, that would be a strong indication that the theory is finite in the perturbation series. If we find a possible one, only explicit calculations of the coefficient in front would tell us if the theory is finite or not to that order.

\end{document}